# To VaR, or Not to VaR, That is the Question

Victor Olkhov

Independent, Moscow, Russia

victor.olkhov@gmail.com

ORCID: 0000-0003-0944-5113

**April 26, 2024**

## ABSTRACT

We consider economic obstacles that limit the reliability and accuracy of value-at-risk (VaR). Investors who manage large market transactions should take into account the impact of the randomness of large trade volumes on predictions of price probability and VaR assessments. We introduce market-based probabilities of price and return that depend on the randomness of market trade values and volumes. Contrary to them, the conventional frequency-based price probability describes the case of constant trade volumes. We derive the dependence of market-based price volatility on the volatilities and correlation of trade values and volumes. In the coming years, that will limit the accuracy of price probability predictions to Gaussian approximations, and even the forecasts of market-based price volatility will be inaccurate and highly uncertain.

This research received no support, specific grants, or financial assistance from funding agencies in the public, commercial, or nonprofit sectors. We welcome valuable offers of grants, support, and positions.

# 1. Introduction

The value-at-risk measure was proposed in the late 1960s, almost 50 years ago, as a response to the request of JP Morgan's Chairman Dennis Weatherstone. "It was of JP Morgan, at the time the Chairman of JP Morgan, who clearly stated the basic question that is the basis for VaR as we know it today – "how much can we lose on our trading portfolio by tomorrow's close?""(Allen, Boudoukh, and Saunders, 2004). The response of JP Morgan's team to Weatherstone's question results in the development of the VaR models by RiskMetrics Group and further studies (Longerstaey and Spencer, 1996; CreditMetrics™, 1997; Duffie and Pan, 1997; Holton, 2002; Allen, Boudoukh, and Saunders, 2004; Choudhry, 2013).

According to Longerstaey and Spencer (1996) "Value-at-Risk is a measure of the maximum potential change in value of a portfolio of financial instruments with a given probability over a pre-set horizon." Since then, Value-at-Risk or VaR, has become a standard tool for risk assessment. As usual, the roots of any good concept like VaR can be found much earlier than it is noted by RiskMetrics "official mythology," and Holton (2002) takes the VaR back to 1922. We cannot refer to all those who contributed to VaR as one of the most effective and useful risk measures and mention only a few (Malkiel, 1981; Marshall and Siegel, 1996; Duffie and Pan, 1997; Berkowitz and O'Brien, 2001; Holton, 2003; Jorion, 2006). Since RiskMetrics publications, the VaR concept has occupied a permanent position in the risk management monographs (Choudhry, 2013; Horcher, 2015). Various forms of the VaR were developed for the risk assessment of market portfolios, corporate risk, credit risk, and financial risk management (Sanders and Manfredo, 1999; Adrian and Brunnermeier, 2011; Andersen et al., 2012). The VaR concept plays an important role in bank and security risk regulations (FRS, 1998; Amato and Remolona, 2005; CESR, 2010). The wide use of VaR as a risk measure is explained by its clear and general concept. Let's take the price probability measure $P(p)$:

$$\int P(p)\ dp = 1 \tag{1.1}$$

and choose a small number $\varepsilon$<<1. Then one can derive the price $p(\varepsilon)$:

$$\int_0^{p(\varepsilon)} P(p)\ dp = \varepsilon \quad ; \quad p(\varepsilon) \le p \ \ with\, probability \ \ 1-\varepsilon \tag{1.2}$$

Price $p(\varepsilon)$ (1.2) determines the bottom line of possible losses with probability $1-\varepsilon$

Simple relations (1.1-1.2) give firm and clear ground for VaR. Only some "easy" problems are left: how to choose and forecast the price probability measure $P(p)$?

In the late 1960s, RiskMetrics developed the first approximations of the VaR. The standard treatment of VaR (Longerstaey and Spencer, 1996) is based on the price probability *P(p)* determined by the number (frequency) of trades at price *p*. To define the price probability *P(p),* one should choose a certain time averaging interval *Δ*, collect all *N* trades with asset *A* during interval *Δ,* and count the number *m(p)* of trades at price *p*. Investors may choose the time interval *Δ* to be equal to an hour, a day, a week, or whatever. The duration of *Δ* impacts the properties of the price probability measure *P(p)*. The frequency-based price probability *P(p)* (1.4) and the n-th statistical moments *p(t;n)* of price during the interval *Δ* (1.3) equal:

$$t-\frac{\Delta}{2}\le t_i\le t+\frac{\Delta}{2} \quad ; \quad i=1,\dots N \tag{1.3}$$

$$P(p)\sim\frac{1}{N}m(p) \quad ; \quad p(t;n)=\frac{1}{N}\sum_k p_k^n\, m(p_k)=\frac{1}{N}\sum_{i=1}^{N}p^n(t_i) \tag{1.4}$$

If one chooses *ε*=5%, then with probability 95% (1.2), all trade prices *p* during interval *Δ* (1.3) will be higher than *p(ε*=5%*).* Hence, *M* shares of asset *A* with a probability 95% will have a value greater or equal than *p(ε*=5%*)M.* Investors may choose the benchmark of 1%, 3%, or whatever and obtain a lower estimate of asset *A* value or possible losses, with a probability 99%, 97%, etc.

As the first approximation, RiskMetrics Group (Longerstaey and Spencer, 1996) assumed that the frequency-based price probability (1.1; 1.4) *P(p)* of trades at price *p* takes the form of standard normal distribution. "A standard property of the normal distribution is that outcomes less than or equal to 1,65 standard deviations below the mean occur only 5 percent of the time" (Longerstaey and Spencer, 1996). Investors have used this result for years as a risk assessment of portfolio losses. Further researchers investigate the way to forecast the frequency-based price probability *P(p)* (1.4), estimate the deviation of price probability *P(p)* (1.4) from the normal distribution, explain the "fat tails" of the observed price probability, etc. These problems are difficult and, till now, far from a final solution.

In this paper we study the problems of the VaR concept: the choice of the price probability and its predictions. We show that the frequency-based price probability (1.4) is not the only one and most likely not the correct one description of random market price. VaR should protect investors from a random change of the market price. We consider randomness of market trade value and volume as the origin of price stochasticity and derive the dependence of market-based statistical moments of price on statistical moments and correlations of the trade values and volumes. That dependence results in a tough conclusion: to predict an average and volatility of price at horizon T, one should predict the averages, volatilities, and correlations of market trade values and volumes at the same horizon T. That nontrivial problem uncovers the economic obstacles that make price probability forecasts rather difficult and uncertain. The distinctions between the market-based and frequency-based price probability (1.4) result in differences in VaR assessments of $p(\varepsilon)$ (1.2; 1.3) and can cause excess losses.

We assume that readers are familiar with statistical moments, etc.

## 2. General considerations

The VaR method is based on predictions of price or return probabilities at horizon *T*. The weaknesses of VaR are hidden in the definition of price probability and the related problems with its predictions. To explain that, we highlight three issues.

The first one states that one can equally describe a random variable by a probability distribution or a set of statistical moments (Shiryaev, 1999; Shreve, 2004). The first *n* statistical moments define the n-approximation of the probability distribution. Predictions of the first *n* statistical moments give predictions of the n-approximation of probability. At the same time, any predictions of probability distribution determine forecasts of statistical moments. The accuracy of the forecasts of the first *n* statistical moments determines the accuracy of probability predictions.

The second one underlines that the price $p(t_i)$ time series is a result of market trades with values $C(t_i)$ and volumes $U(t_i)$. One can describe a market trade at time $t_i$ by its value $C(t_i)$, volume $U(t_i)$, and price $p(t_i)$:

$$C(t_i) = p(t_i)U(t_i) \tag{2.1}$$

We consider the trade value $C(t_i)$, volume $U(t_i)$, and price $p(t_i)$ (2.1) during the averaging interval $\Delta$ (3.1) as random variables. If random values $C(t_i)$, volumes $U(t_i)$, and prices $p(t_i)$ satisfy the equation (2.1), then it is independently impossible to define their statistical moments or their probability distributions. We describe the dependence of market-based average and volatility of prices on the averages, volatilities, and correlations of trade values and volumes. That gives the Gaussian approximations of market price probability.

Our third issue concerns the predictions of price statistical moments at horizon *T*. We argue that to forecast at horizon T the market-based average and volatility of price, one should predict averages, volatilities, and correlations of market trade values and volumes at the same horizon *T*. We discuss the economic obstacles that prevent predictions of the 2nd and higher statistical moments of market trade. In the best case, in the coming years, the predictions of trade statistical moments and, hence, predictions of price statistical moments, will be limited by predictions of the 2nd statistical moments. Hence, predictions of price probability will be limited to Gaussian distributions only. Below, we consider these issues in more detail.

In the next section, we introduce the market-based average and volatility of a random price (Olkhov, 2021a; 2022; 2023a) that define a Gaussian approximation of the price probability and discuss distinctions between the market-based and frequency-based statistical moments (1.4). Further, we discuss the economic obstacles that prevent precise prediction of market-based price volatility, which make the use of VaR rather uncertain. One can find a description of market-based statistical moments of return in Olkhov (2023a; 2023b).

## 3. Market-based average and volatility of price

Let us assume that the time interval ε between trades (2.1) is rather small and constant. High-frequency trade time series behave irregularly or randomly during almost any averaging interval $\Delta$. The choice of the averaging interval $\Delta$ allows estimate random properties of the trade value $C(t_i)$, volume $U(t_i)$, and price $p(t_i)$ during $\Delta$ and then make an attempt to predict them at horizon $T$. For convenience, we assume that times $t_i$ belong to the averaging interval $\Delta$ (3.1) near moment $t$ if:

$$t-\frac{\Delta}{2}\leq t_i\leq t+\frac{\Delta}{2}\ ;\ \ t_i=t_0+(i-1)\cdot\varepsilon\ ;\ \ i=1,\ldots N \tag{3.1}$$

The duration of the averaging interval $\Delta$ defines the number of members of the time series of the trade values $C(t_i)$ and volumes $U(t_i)$. The *n-th* statistical moments of trade value $C(t;n)$ and volume $U(t;n)$ take the form:

$$C(t;n)=E[C^n(t_i)]\sim\frac{1}{N}\sum_{i=1}^{N}C^n(t_i)\quad ;\quad U(t;n)=E[U^n(t_i)]\sim\frac{1}{N}\sum_{i=1}^{N}U^n(t_i) \tag{3.2}$$

We use the symbol "~" in (3.2) to highlight that a finite number $N$ of terms defines the estimate of the *n-th* statistical moments of the trade value and volume. More than 35 years ago, Berkowitz et al. (1988) introduced the volume weighted average price (VWAP), which is widely used now (Buryak and Guo, 2014; Duffie and Dworczak, 2018; CME Group, 2020). The VWAP $p(t;1,1)$ or price 1-st statistical moment determined by trade values $C(t_i)$ and volumes $U(t_i)$ during the interval $\Delta$ (3.1) takes the form:

$$p(t;1,1)=\frac{1}{U_\Delta(t;1)}\sum_{i=1}^{N}p(t_i)\,U(t_i)=\frac{1}{U_\Delta(t;1)}\sum_{i=1}^{N}C(t_i)=\frac{C_\Delta(t;1)}{U_\Delta(t;1)}=\frac{C(t;1)}{U(t;1)} \tag{3.3}$$

$$C_\Delta(t;n)=NC(t;n)=\sum_{i=1}^{N}C^n(t_i)\quad ;\quad U_\Delta(t;n)=NU(t;n)=\sum_{i=1}^{N}U^n(t_i) \tag{3.4}$$

The 1st statistical moments $C(t;1)$ and $U(t;1)$ (3.2) denote the average value and volume of $N$ trades during $\Delta$ (3.1). The VWAP price $p(t;1,1)$ or 1-st statistical moment of price is determined as ratio of the mean value *C(t;1)* to mean volume *U(t;1)* (3.3) or as ratio of total value $C_\Delta(t;1)$ to total volume $U_\Delta(t;1)$ (3.4) during $\Delta$ (3.1). Let us transfer (3.3) as follows:

$$p(t;1,1) = \frac{1}{U_\Delta(t;1)} \sum_{i=1}^{N} p(t_i)\,U(t_i) = \sum_{i=1}^{N} p(t_i)\,w(t_i;t,1) \quad (3.5)$$

$$w(t_i;t,1) = \frac{U(t_i)}{\sum_{i=1}^{N} U(t_i)} = \frac{U(t_i)}{U_\Delta(t;1)} \quad ; \quad \sum_{i=1}^{N} w(t_i;t,1) = 1 \quad (3.6)$$

We highlight that the functions $w(t_i;t,1)$ (3.5; 3.6) have meaning of weight functions, but don't play role of price probabilities. Let us take the n-th degree of (2.1):

$$C^n(t_i) = p^n(t_i)\,U^n(t_i) \quad (3.7)$$

We define price statistical moments *p(t;m,n)* (3.8) as the m-th degree of price $p^m(t_i)$ averaged over the *n-th* weight functions $w(t_i;t,n)$ (3.9) that have form similar to VWAP (3.3; 3.5):

$$p(t;m,n) = \sum_{i=1}^{N} p^m(t_i) w(t_i;t,n) = \frac{1}{\sum_{i=1}^{N} U^n(t_i)} \sum_{i=1}^{N} p^m(t_i) U^n(t_i) \quad (3.8)$$

$$w(t_i;t,n) = \frac{U^n(t_i)}{\sum_{i=1}^{N} U^n(t_i)} = \frac{U^n(t_i)}{U_\Delta(t;n)} \quad ; \quad \sum_{i=1}^{N} w(t_i;t,n) = 1 \quad (3.9)$$

Relations (3.8) define a set of price statistical moments averaged over different weight functions $w(t_i;t,n)$. We use the set of price statistical moments (3.8) to define market-based statistical moments of price. For each *n=1,2,…*, the statistical moments of price *p(t;m,n)* (3.8) make their contribution into the *n-th* market-based statistical moment of price *a(t;n)*:

$$a(t;n) = E_m[p^n(t_i)] \quad ; \quad n = 1,2,3,\ldots \quad (3.10)$$

We denote market-based mathematical expectation $E_m[..]$ to distinguish it from frequency-based mathematical expectation (3.2). As market-based average or the 1st statistical moment *a(t;1)* of price we take VWAP *p(t;1,1)* (3.3):

$$a(t;1) = E_m[p(t_i)] = p(t;1,1) \quad (3.11)$$

To define the 2nd price statistical moment *a(t;2)* and price volatility $\sigma^2(t)$ (3.12):

$$a(t;2) = E_m[p^2(t_i)] \quad ; \quad \sigma^2(t) = E_m\left[\big(p(t_i) - a(t;1)\big)^2\right] = a(t;2) - a^2(t;1) \quad (3.12)$$

one should reconcile the market-based average price *a(t;1)* (3.11) with the price statistical moments *p(t;m,2)* determined by the weight functions $w(t_i;t,2)$ (3.9). In particular, it is important to prove that price volatility (3.12) would always be non-negative $\sigma^2(t) \geq 0$. To get that, we define the market-based price volatility $\sigma^2(t)$ as follows:

$$\sigma^2(t) = \sum_{i=1}^{N} \big(p(t_i) - a(t;1)\big)^2 w(t_i;t,n) = p(t;2,2) - 2p(t;1,2)a(t;1) + a^2(t;1) \geq 0 \quad (3.13)$$

One can present relations (3.12; 3.13) as follows (Olkhov, 2021a; 2022):

$$\sigma^2(t) = \frac{\Omega_C^2(t) + a^2(t;1)\Omega_U^2(t) - 2a(t;1)corr\{C(t)U(t)\}}{U(t;2)} \quad (3.14)$$

$$a(t;2) = \frac{C(t;2) + 2a^2(t;1)\Omega_U^2(t) - 2a(t;1)corr\{C(t)U(t)\}}{U(t;2)} \quad (3.15)$$

Price volatility $\sigma^2(t)$ (3.14) and the 2nd statistical moment of price $a(t;2)$ (3.15) depend on volatilities of market trade values $\Omega_C^{\,2}(t)$ and volumes $\Omega_U^{\,2}(t)$ (3.16):

$$\Omega_C^2(t) = C(t;2) - C^2(t;1) \quad ; \quad \Omega_U^2(t) = U(t;2) - U^2(t;1) \tag{3.16}$$

and on correlation $corr\{C(t)U(t)\}$ (3.17) between trade values and volumes.

$$corr\{C(t)U(t)\} = E[C(t_i)U(t_i)] - C(t;1)U(t;1) \tag{3.17}$$

The joint average $E[C(t_i)U(t_i)]$ (3.18) of the product of trade value and volume takes the form:

$$E[C(t_i)U(t_i)] \sim \frac{1}{N}\sum_{i=1}^{N} C(t_i)U(t_i) \tag{3.18}$$

Relations (3.11; 3.14-3.18) determine the dependence of the market-based average and volatility of price on averages, volatilities, and correlations of trade values and volumes. The derivation of the dependence of the 3rd and 4th statistical moments of price on statistical moments of trade values and volumes is given in Olkhov (2022). The first two market-based price statistical moments – average $a(t;1)$ (3.11) and volatility $\sigma^2(t)$ (3.14) - determine the Gaussian approximation of price probability. It is obvious that huge amounts of market trade records permit derive higher market-based price statistical moments and describe a more precise market price probability "today." However, as we show below, that does not help to predict a more precise price probability at horizon *T*. In many years to come, the Gaussian approximation of price probability will remain the only approximation of the predicted price probability. In the next section, we consider economic obstacles that limit the forecasting of the market-based price probability by Gaussian approximations. That definitely limits the reliability of VaR use.

## 4. Economic obstacles that limit the accuracy of probability predictions

At first, let us compare the frequency-based price statistical moments $p(t;n)$ (1.4) and statistical moments $p(t;m.n)$ (3.8) that are determined by the weight functions $w(t_i;t,n)$ (3.9). If all trade volumes $U(t_i)$ are constant during the averaging interval $\Delta$ (3.1), then:

$$p(t;n) = \frac{1}{N}\sum_{i=1}^{N} p^n(t_i) = p(t;n,m) \;\; ; \;\; n,\; m = 1,2,.. \tag{4.1}$$

In this case, the price volatility $\sigma^2(t)$ (3.14) takes the form of frequency-based price volatility:

$$\sigma^2(t) = p(t;2) - p^2(t;1) \tag{4.2}$$

The main contribution of the introduction of market-based volatility $\sigma^2(t)$ (3.14) is that relations (3.14) describe the impact of the randomness of trade volumes on price volatility. The volatilities of market trade value $\Omega_C^{\,2}(t)$ and volume $\Omega_U^{\,2}(t)$ (3.16) and their correlation $corr\{C(t)U(t)\}$ (3.17) determine market-based price volatility. That is the main distinction between the price volatility $\sigma^2(t)$ (3.14) and the frequency-based price volatility (4.2), which gives an estimate in the case that all trade volumes are constant during the averaging interval (3.1). We repeat this statement as it uncovers hidden economic obstacles that make the predictions of the market-based price volatility $\sigma^2(t)$ (3.14) a rather complex problem.

Market participants who make market trades with large volumes should consider the impact of the randomness of trade volumes on price volatility $\sigma^2(t)$ (3.14). The use of frequency-based price volatility (4.2) has mostly psychological effects as it doesn't describe the impact of random market trade with large volumes. The differences between market-based volatility (3.14) and frequency-based volatility (4.2), which determine Gaussian approximations and corresponding VaR assessments (1.2), could result in excess losses.

However, the accuracy of predictions of market-based price volatility (3.14) is limited by internal economic obstacles. Let us briefly consider two main issues. The first one relates to the fact that price volatility $\sigma^2(t)$ (3.14) predictions at horizon *T* depend on forecasting of market trade value $\Omega_C^2(t)$ and volume $\Omega_U^2(t)$ (3.16) volatilities and their correlation $corr\{C(t)U(t)\}$ (3.17). In simple words, to predict market-based price volatility (3.14), one should predict the properties of random market trade at horizon *T*. If one accidentally succeeds in accurate forecasting averages, volatilities, and correlations of trade values and volumes at horizon T that can be equal to a day, a week, a month, or whatever, this lucky one will be able to manage his market trades with much more personal benefits than projecting a Gaussian approximation of price probability for VaR. But some economic obstacles make such accurate forecast rather uncertain. That is the second issue we highlight.

Indeed, the description of the 2nd statistical moments of trade values $C(t;2)$ and volume $U(t;2)$ (3.2) introduces the new class of macroeconomic variables that we call the 2nd degree variables. They are composed of sums of the 2nd degree of market trade values, or volumes. Almost all macroeconomic variables that are used now to describe economic evolution are composed of the sums of the 1st degrees of market transactions (Fox et al., 2017). Macroeconomic investment, credits, and consumption are sums of the corresponding variables of economic agents. In turn, agents' variables are composed of the sums of their market trades during the interval $\Delta$ (3.1) – investment, credit, and consumption. We call the description of the 1st degree macroeconomic variables 1st-order economic theory. However, agents make their transactions under their economic expectations, which could be formed by forecasts of price and return averages and volatilities. Volatilities of market trade values and volumes, of price, return, and volatilities of other economic variables such as demand and supply, for example, are the origin of economic variables of the 2nd degree that significantly impact agents' trade decisions and thus impact economic evolution and sustainability. The description of the 2nd degree macroeconomic variables needs the development of the economic theory of the 2nd order that is absent now. One should develop a methodology for collecting the 2nd degree pairs for almost all economic variables of the 1st degree. The methodology similar to (Fox et al., 2017) should govern econometric data to define the 2nd degree investments, credits, or consumptions composed of squares of corresponding trade values and volumes. To describe the set of 1st and 2nd degree economic variables, one should develop economic theories of the 2nd order. Actually, the attempts to predict price and return probabilities with higher accuracy than Gaussian distributions will create a need for a description of the 3rd and 4th statistical moments of price and return. That in turn will require a description of the 3rd and 4th statistical moments of market trade values and volumes (3.2), determined by the sums of the 3rd and 4th degrees of trade values and volumes. Thus the complexity of economic modeling will grow with the attempts to predict price and return probabilities with greater accuracy.

In the coming years, the accuracy of the forecasts of price and return probabilities will be limited by Gaussian distributions. The development of econometric methodology and economic theories of the 2nd order and higher that would describe the evolution of 1st and 2nd degree macroeconomic variables is a problem for the long-term future (Olkhov, 2021b; 2023c; 2023d). Till then, even the predictions of price volatility will remain inaccurate and uncertain. That essentially limits the reliability of VaR.

## 5. Conclusion

The forecasts of price and return statistical moments or probabilities are at the heart advanced economic and financial studies. After 50 years of use, the main problems with the VaR concept are still open. Only the development of econometric methodology that can govern the collection and verification of macroeconomic variables of the 1st and 2nd degrees and the creation of macroeconomic theories of the 2nd order that will describe the mutual evolution of these variables could significantly improve the predictions of price and return volatilities. Till then, any such predictions remain highly uncertain and may be more harmful than useful for investors.

## References

Adrian, T., and M. K. Brunnermeier, (2011). COVAR. NBER, Cambridge, WP 17454, 1-45

Allen, L., Boudoukh, J., and A. Saunders, (2004). Understanding Market, Credit, And Operational Risk. The Value At Risk Approach, Blackwell Publ., Oxford, UK. 1-313

Amato, J.D., and E. M. Remolona, (2005). The Pricing of Unexpected Credit Losses, BIS WP 190, 1-46

Andersen, T.G., Bollerslev, T., Christoffersen, P.F., and F. X. Diebold, (2012). Financial Risk Measurement For Financial Risk Management. NBER, Cambridge, WP 18084, 1-130

Berkowitz, J., and J. O'Brien, (2001). How Accurate are Value-at-Risk Models at Commercial Banks? Fed. Reserve Board, Washington, D.C. 1-28

Berkowitz, S.A., Dennis E., Logue, D.E., and E. A. Noser, Jr., (1988). The Total Cost of Transactions on the NYSE, *The Journal Of Finance*, 43, (1), 97-112

Buryak, A. and I. Guo, (2014). Effective And Simple VWAP Options Pricing Model, *Intern. J. Theor. Applied Finance*, 17, (6), 1450036, https://doi.org/10.1142/S0219024914500356

CESR, (2010). CESR's Guidelines on Risk Measurement and the Calculation of Global Exposure and Counterparty Risk for UCITS. Comm. of Euro. Securities Regulators, 1-43

Choudhry, M., (2013). An Introduction to Value-at-Risk, 5th Edition. Wiley, 1-224

CME Group, (2020). www.cmegroup.com/confluence/display/EPICSANDBOX/GovPX+Historical+Data ; www.cmegroup.com/confluence/display/EPICSANDBOX/Standard+and+Poors+500+Futures

CreditMetrics™, (1997). Technical Document. J.P. Morgan & Co, NY. 1-212

FRS, (1998). Trading and Capital-Markets Activities Manual. Division of Banking Supervision and Regulation, Board of Governors of the FRS, Washington DC. 1-672

Fox, D.R. et al. (2017). Concepts and Methods of the U.S. National Income and Product Accounts. BEA, US.Dep. Commerce, 1-447

Duffie, D., and J. Pan, (1997). An Overview of Value-at-Risk, *J. of Derivatives*, 4, (3), 7-49

Duffie, D. and P. Dworczak, (2018). Robust Benchmark Design, NBER WP 20540, 1-56

Holton, G. A., (2002). History of Value-at-Risk: 1922-1998. Working Paper. Contingency Analysis, Boston, MA. 1-27

Holton, G. A., (2003). Value-at-Risk: Theory and Practice, San Diego: Academ. Press. 1-405

Horcher, K. A., (2015). Essentials of financial risk management. John Wiley&Sons, Inc., New Jersey. 1-272

Jorion, P, (2006). Value At Risk: The New Benchmark For Managing Financial Risk, 3-d Ed., McGraw-Hill.

Longerstaey, J., and M. Spencer, (1996). RiskMetrics -Technical Document. J.P.Morgan &

Reuters, N.Y., Fourth Edition, 1-296
Malkiel, B.G., (1981). Risk And Return: A New Look. NBER, Cambridge, WP 700, 1-45
Marshall, C., and M. Siegel, (1996). Value at Risk: Implementing a Risk Measurement Standard, The Wharton Financial Institutions Center, 1-34
Olkhov, V., (2021a). Three Remarks On Asset Pricing. MPRA WP119863, 1-20
Olkhov, V., (2021b). Theoretical Economics and the Second-Order Economic Theory. What is it?, MPRA WP 120536, 1-13
Olkhov, V. (2022). Market-Based Asset Price Probability, MPRA WP120026, 1-18
Olkhov, V., (2023a). Market-Based Probability of Stock Returns, SSRN WP 4350975, 1-17
Olkhov, V., (2023b). Market-Based 'Actual' Returns of Investors, MPRA WP 120285, 1-16
Olkhov, V., (2023c). Economic Complexity Limits Accuracy of Price Probability Predictions by Gaussian Distributions, SSRN WP 4550635, 1-23
Olkhov, V., (2023d). Theoretical Economics as Successive Approximations of Statistical Moments, SSRN WP 4586945, 1-17
Sanders, D. R., and M. R. Manfredo, (1999). Corporate Risk Management and the Role of Value-at-Risk. Proc. NCR-134 Conf. on Applied Commodity Price Analysis, Forecasting, and Market Risk Management. Chicago, IL. 1-14
Shiryaev, A.N., (1999). Essentials Of Stochastic Finance: Facts, Models, Theory. World Sc. Pub., Singapore. 1-852
Shreve, S. E. (2004). Stochastic calculus for finance, Springer finance series, NY, USA